\documentclass[aps,twocolumn,showpacs,superscriptaddress,citeautoscript,prb,floatfix]{revtex4-2}

\usepackage{graphicx} 
\usepackage{amsmath, amssymb, mathrsfs}
\usepackage{amsfonts}
\usepackage{tabularx}
\usepackage{cancel}
\usepackage{ulem}
\usepackage{bbold} 
\usepackage{wrapfig}

\usepackage{bm}
\usepackage[T1]{fontenc}
\usepackage{textcomp}
\usepackage{enumitem}
\usepackage{cancel}
\usepackage{bbold}
\usepackage{esvect}
\usepackage{commath}
\usepackage[most]{tcolorbox}
\usepackage{lipsum, babel}
\usepackage{comment}
\usepackage{lineno}
\usepackage{physics}
\usepackage{xcolor}
\usepackage{hyperref}
\usepackage{adjustbox}

\newcommand{\angstr}{\textup{~\AA}}

\usepackage[left=23mm,right=23mm, top = 20mm, bottom = 20mm, paper = a4paper]{geometry}

\begin{document}




\title{Spin-dependent transport through edge states in 2D semi-Dirac materials with Rashba spin-orbit coupling and band inversion}

\author{Marta Garc\'{\i}a Olmos}
\email{mgarcia.o@usal.es}
\affiliation{%
  Nanotechnology Group, USAL—Nanolab and IUFFyM, University of Salamanca\\
 Plaza de la Merced, Edificio Triling\"{u}e, 37008, Salamanca, Spain.
}%

\affiliation{Instituto de Estructura de la Materia IEM-CSIC, Serrano 123, E-28006 Madrid, Spain}

\author{Yuriko Baba}
\email{yuriko.baba@uam.es}

\affiliation{
Department of Theoretical Condensed Matter Physics, Condensed
Matter Physics Center (IFIMAC)
Universidad Autónoma de Madrid, E-28049 Madrid, Spain}

\affiliation{GISC, Departamento de F\'{\i}sica de Materiales, Universidad Complutense, E-28040 Madrid, Spain}

\author{Alexander López}
\email{alexlop@espol.edu.ec}

\affiliation{GISC, Departamento de F\'{\i}sica de Materiales, Universidad Complutense, E-28040 Madrid, Spain}

\affiliation{%
 Escuela Superior Politécnica del Litoral, ESPOL, Departamento de Física, Campus Gustavo Galindo Km. 30.5 vía perimetral, P.O. Box 09-01-5863, Guayaquil, Ecuador.
}

\author{Mario Amado}
\email{mario.amado@usal.es}
\affiliation{%
  Nanotechnology Group, USAL—Nanolab and IUFFyM, University of Salamanca\\
 Plaza de la Merced, Edificio Triling\"{u}e, 37008, Salamanca, Spain.
}%

\author{Rafael A. Molina}
\email{rafael.molina@csic.es}
\affiliation{Instituto de Estructura de la Materia IEM-CSIC, Serrano 123, E-28006 Madrid, Spain}

\
\begin{abstract}
We investigate the bulk-boundary correspondence in two-dimensional type-I semi-Dirac materials with band inversion and Rashba spin-orbit coupling. Employing a dimensional reduction framework, we identify the Zak phase along the quadratically dispersing direction as a topological invariant that captures the presence of edge states. In the non-trivial topological regime, systems with finite width exhibit energy-dependent edge states that are topologically protected only at specific momenta. At $k_x=0$, symmetry-protected edge states emerge, analogous to the Rashba-free case. At finite $k_x$, the interplay of spin-orbit coupling and band structure gives rise to spin-dependent edge states, localized on specific edges based on its spin and particle-hole character.  We compute spin-resolved conductance through these edge channels and observe robust, tunable oscillations—attributable to spin precession induced by the effective Rashba magnetic field. These results reveal how spin-orbit interactions enrich the edge physics of semi-Dirac systems and provide a platform for spintronic control in anisotropic topological materials.

\end{abstract}

\maketitle

\section{Introduction}

Spin-orbit coupling (SOC) is a relativistic interaction that couples the electron's spin to its orbital motion, lifting spin degeneracies, modifying the band structures and the electronic response of materials. In low-dimensional systems, SOC is particularly crucial, since reduced symmetry and quantum confinement of electrons can significantly enhance its effects \cite{Ahn2020}. These conditions make SOC a key ingredient in the exploration of next-generation nanoelectronic devices, from enabling spin-dependent transport in nanostructures \cite{reviewZutic2004} to the emergence of nontrivial topological phases \cite{Manchon2015, Zollner2025}.

One particularly attractive form of the SOC is the Rashba interaction (RSOC), which arises in systems lacking structural inversion symmetry \cite{Bychkov1984}. The RSOC typically originates from an out-of-plane electric field that, in the electron's rest frame, manifests as a momentum-dependent magnetic field lying in the plane of motion. This field couples to the electron spin and splits the spin-degenerate bands. This type of SOC must be distinguished from others, such as the Dresselhaus SOC, which arises from bulk inversion asymmetry in the crystal lattice \cite{Dresselhaus1955}. While the latter is intrinsic to the material, the Rashba one is highly tunable and can be externally modulated by electric fields or chemical doping, making it extremely relevant for technological applications. 
Building on this, the Datta and Das transistor proposed in the early nineties introduced the possibility to control the spin via the Rashba effect. This device opened the door to manipulate spin-polarized currents without external magnetic fields \cite{Datta_Das_transist}. Experimental demonstrations of this concept have been achieved in InGaAs/InAlAs heterostructures \cite{Nitta1997} and HgTe quantum wells \cite{Schultz1999}.

\par In this work, we explore the role of RSOC in materials exhibiting a semi-Dirac dispersion relation, an exciting electronic structure in which quasiparticles behave as massless Dirac fermions along a specific  direction of motion and as massive particles along the perpendicular one. Therefore, the low-energy spectrum is linear in one momentum direction and quadratic in the orthogonal axis, leading to unconventional transport and optical responses \cite{Carbotte2019, Huang2023, Link2018, Xiong2023, Chan2023}. The theoretical proposal of semi-Dirac dispersion arose from studies of uniaxially strained graphene \cite{Dietl2008, Montambaux2009, Delplace2010}. However, the extremely large strains required (on the order of 20\%) to merge the Dirac cones make this realization experimentally challenging \cite{Goerbig2008}. Beyond strained graphene, semi-Dirac dispersion has been explored in a wide variety of systems, such as TiO$_2$/VO$_2$ nanostructures \cite{Pardo2009, Banerjee2009, Huang2015}, organic salts \cite{Katayama2006}, modified silicene \cite{Zhong2016}, few layers of black phosphorus \cite{CastellanosGomez2014, Rodin2014, PeraltaMireles2024}, surface states of three-dimensional bulk  topological insulators \cite{Zhai2011, Li2011} or semimetals \cite{Smith2024}, and even synthetic materials like engineered photonic or cold-atom lattices \cite{Wu2014, PhysRevX.9.031010, Bellec2013}, and polariton honeycomb lattices \cite{Real2020}, where the hopping parameters can be precisely tuned.

Building on these diverse platforms, different works have investigated the possibility of engineering non-trivial topology by adding symmetry-allowed terms, i.e., mass terms \cite{Uryszek2019, Kotov2021}, high-order hoppings \cite{Mohanta2021, Mondal2022}, spin-orbit interaction \cite{Huang2015}, or periodic drives \cite{Saha2016}, which can invert the character of the bands at the semi-Dirac point. To organize this rich landscape, semi-Dirac models are often classified into three types on the basis of the structure of their low-energy Hamiltonians. The type-I Hamiltonians are separable in momentum components and typically lack cross terms. Type-II semi-Dirac low-energy Hamiltonians exhibit non-separable momentum terms, such as $k_xk_y$, leading to the merging of three Dirac cones into a semi-Dirac point. When intrinsic SOC is present, these systems can become a Chern insulator with $C = -2$ \cite{Huang2015}. Similar dispersions have experimentally been identified in the topological metal $\mathrm{ZrSiS}$, where low-energy semi-Dirac excitations merge in nodal lines protected by crystalline symmetry \cite{PRX2024}. 
Type-III systems combine a flat band with a dispersive one, producing a linear-flat dispersion at a critical point. These have been proposed in striped boron sheets \cite{Zhang2017} and have been realized in photonic lattices under synthetic strain that exhibit a non-trivial topological charge \cite{PhysRevX.9.031010}.\\

Despite the extensive literature, most of the existing studies have focused on bulk properties, leaving the behavior of edge states, especially under the influence of SOC, largely unexplored. Recent \textit{ab initio} results for black phosphorus suggest that SOC can have a significant impact even in type I-like systems \cite{PeraltaMireles2024}, highlighting the need to consider spin-dependent effects.
Our work is grounded in the type-I semi-Dirac framework but considers an extended model where additional symmetry-allowed mass terms are introduced following ref. \cite{SDpaper_2024}. These additional terms preserve the semi-Dirac nature, but can induce band inversion and lead to emergent topological features. Although such systems do not exhibit full two-dimensional topological protection due to the zero total Chern number $(C = 0)$, they can host momentum-dependent topological states characterized by a non-trivial Zak phase along specific directions. By incorporating spin-flipping processes due to RSOC into this framework, our goal is to explore their impact on the topological and transport properties of the semi-Dirac material.\\

The paper is organized as follows. In section \ref{sec:Rashba-effect}, we introduce the Rashba spin-orbit interaction in our particular mode and analyze its effect on both the bulk and edge states dispersion. This section is divided into three parts: first, we derive the corresponding RSOC Hamiltonian (sec. \ref{sec:Deriving_Rashba}); next , we  study how the RSOC term splits the energy bands and modifies the spin texture of the edge states in a semi-infinite geometry (sec. \ref{sec:Dispersion_edge_states}). The edge states' dispersion in the presence of Rashba coupling is obtained using perturbation theory. Finally, we analyze the RSOC effect on the topological features (sec. \ref{sec:Topological_protect}).  We distinguish states at $k_x = 0$ and states at $k_x \neq 0$ as two different kinds of non-trivial points. In section \ref{sec:spin_resolvedG}, we turn to transport properties. Numerically, we analyze the spin-resolved conductance in a two-terminal device using a tight-binding version of our Hamiltonian within the Landauer-Büttiker formalism. We find that spin precession induces a large value of the spin-flip conductance, which oscillates with both the system length and the strength of the Rashba coupling. 
The robustness of these effects is tested under different types of disorder. Our findings provide insight into the potential for engineering spin-dependent edge phenomena in anisotropic systems without full topological protection.

\section{General properties of Rashba spin-orbit coupling in semi-Dirac materials}  \label{sec:Rashba-effect}

Our starting point is the low-energy Hamiltonian of the semi-Dirac type-I system,
$H_0 (\boldsymbol{k}) = \boldsymbol{d}(\boldsymbol{k}) \cdot \boldsymbol{\sigma} (\boldsymbol{k})$, where
\begin{equation} \label{eq:d-vector}
    \boldsymbol{d}(\boldsymbol{k}) = \begin{pmatrix} V_x k_x^2 \\ V_y k_y \\ M_0 - M_{1x}k_x^2 - M_{1y} k_y^2 \end{pmatrix}.
\end{equation}
Here, $\boldsymbol{\sigma} = (\sigma_x, \sigma_y, \sigma_z)$ are the Pauli matrices that act on the sublattice space, and $V_x, V_y, M_0, M_{1x}$ and $M_{1y}$ are the model parameters. This Hamiltonian considers a single spin polarization, effectively ignoring spin effects. 

\par The associated energy spectrum is gapped for generic values of $M_0$. However, at $M_0 = 0$, the gap closes at $(k_x, k_y) = (0,0)$, indicating a critical point between distinct insulating phases. In particular, when $\operatorname{sgn}(M_0/M_{1y})>0$, the gap reopens in a band-inverted regime that can host edge states. Due to the anisotropy of the model, these edge states appear only when a system with open boundary conditions in the $y$-direction is considered, the direction on which the Hamiltonian depends linearly on the momentum. In contrast, a finite-size system in the $x$-direction does not support edge states.

\par Unlike in Dirac systems that display a linear dispersion relation, in the semi-Dirac scenario the edge states dispersion turns out to be quadratic along the $k_x$ direction of the momentum space, $E_{\pm} = \pm V_x \text{sgn}(M_{1y} V_y) k_x^2$. The $\pm$ subindex is linked to the sign of the energy band but also to the location of the state, $+$ for the upper edge and $-$ for the lower edge. In real space, the edge states wavefunction is given by,
\begin{equation} \label{eq:edge _states_wf}
    \psi_{\pm} (k_x, y) \sim e^{ik_x x} \left[ e^{\lambda_1 (\pm y - W/2)} - e^{\lambda_2 (\pm y - W/2)}\right] \phi_{\pm} ,
\end{equation}
with
\begin{align}
    & \phi_{\pm} = \frac{1}{\sqrt{2}} \begin{pmatrix} 1 \\ \pm \text{sgn} (M_{1y}V_y)\end{pmatrix}, \\
    & \lambda_{1,2} = \frac{V_y}{M_{1y}} \pm \frac{1}{M_{1y}} \sqrt{V_y^2 -M_0 M_{1y} + M_{1x} M_{1y}k_x^2}.
\end{align}
The case with $k_x =0$, is particularly relevant, since the Hamiltonian reduces to an effective one-dimensional model with linear bulk dispersion as a function of $k_y$. In this situation, the system can be mapped onto a generalized Su-Schrieffer-Heeger (SSH) model by applying a rotation $R = \exp (i \sigma_y \pi/4)$ in the Pauli matrix space \cite{SDpaper_2024}.  Then, the topological properties of the states are characterized by a non-trivial Zak phase of $\pi $ \cite{Vanderbilt_2018}. As a result, despite the absence of a fully two-dimensional topological invariant, the system hosts a momentum-dependent topological edge state consistent with the bulk-boundary correspondence. Although additional non-protected edge states may be present, the existence of a single symmetry-protected edge mode at $k_x = 0$ highlights the topological character of the system. We will formally describe this in sec. \ref{sec:Topogical_protect}.\\

To generalize the model, we incorporate the spin degree of freedom by considering the spin states as time-reversal partners, similar to the case of the spin Hall effect model \cite{QSHEreview}. Yet, in contrast to that case, the spin-polarized Hamiltonian given by \eqref{eq:d-vector} already preserves time-reversal symmetry. Therefore,  
the extended model is formed by two identical degenerate copies, one for each spin orientation. Choosing $z$ as the quantization axis and using the basis $\left\lbrace \left| \uparrow \right\rangle, \left| \downarrow \right\rangle \right\rbrace \otimes \left\lbrace \left| A \right\rangle, \left| B \right\rangle \right\rbrace$, where $\ket{A}$ and $\ket{B}$ denote the sublattice basis, the Hamiltonian is block-diagonal in absence of any other interaction, 
\begin{equation} \label{eq:ham_SD}
    H_{\rm SD}^{4\times4} (\boldsymbol{k}) = \begin{pmatrix} H_0 & 0_{2\times 2} \\ 0_{2\times 2} & H_0 
    \end{pmatrix} .
\end{equation}
So far we have summarized the main properties of the two-band model and we have introduced the spin degree of freedom into our framework. 

\subsection{Derivation of the Rashba term} \label{sec:Deriving_Rashba}

We will now consider a Rashba-like interaction between the spins. SOC interaction arises because of the coupling between spin degrees of freedom and the electron's motion, which is described by the velocity operator. It takes the general form $H_{\rm SOC} \sim (\boldsymbol{s} \times \boldsymbol{v}) \cdot \boldsymbol{E}$, where $\boldsymbol{s}$ is the spin operator and $\boldsymbol{v}$ is the velocity operator \cite{Winkler2003}. As we mentioned above, the RSOC comes from the presence of an electric field that induces an effective magnetic field in the moving charge carriers. That magnetic field is oriented perpendicular to both the motion direction and the applied field.\\

To be specific, we consider a sample in the $xy$-plane and an out-plane electric field, $\boldsymbol{E} = E_z \hat{z}$. The velocity operator is given by the commutator between the position operator and the Hamiltonian $\hat{v} = 2 (V_x \sigma_x - M_{1x} \sigma_z) k_x\hat{x} + (V_y \sigma_y - 2 M_{1y} k_y \sigma_z) \hat{y}$. Working in the basis $\lbrace | A \uparrow \rangle, | B \uparrow \rangle, | A \downarrow \rangle, | B \downarrow \rangle \rbrace$ the full Hamiltonian is a $4 \times 4$ matrix, where the RSOC appears exclusively in the $2\times 2$ off-diagonal blocks that connect spin-up and spin-down sectors,
\begin{widetext}
\begin{subequations}\label{eq:H_RSOC}
    \begin{equation}
      H_{\rm RSOC} = H_{\rm SD}^{4 \times 4} + \alpha U_{\rm RSOC}~,  
    \end{equation}
    where
    \begin{align}
        H_{\rm SD}^{4 \times 4} & = (M_0 - M_{1x} k_x^2 - M_{1y} k_y^2) (s_0 \otimes \sigma_z) + (V_x k_x^2) (s_0 \otimes \sigma_x) + V_y  k_y (s_0 \otimes \sigma_y)~, \\
        U_{\rm RSOC} & =   V_x k_x (s_y \otimes \sigma_x) - M_{1x} k_x(s_y \otimes \sigma_z) + M_{1y} k_y (s_x \otimes \sigma_z) -
     (V_y/2)(s_x \otimes \sigma_y)~. 
     \label{eq:H_RSOC:U}
    \end{align}
\end{subequations}
\begin{table}[h] 
    \begin{center}
        \begin{tabular}{| c | c | c | c | c | c |} 
        \hline 
        $M_0 \left[ \text{eV} \right]$  & $M_{1x} [ \text{eV} \cdot \text{ \AA}^2 ]$ &$M_{1y} [ \text{eV} \cdot \text{ \AA}^2 ]$ &$V_x [ \text{eV}\cdot \text{ \AA}^2 ]$ & $V_y [ \text{eV}\cdot \text{ \AA} ]$ & $\alpha [ \text{eV} \cdot \text{s} \cdot \text{\AA}^{-1} ]$ \\ \hline
        0.09 & 0.23 & 0.23 & -0.38 & -0.5 & 0.2 \\ \hline
        \end{tabular}
        \caption{Set of model parameters chosen for the present figures. $M_0$ is a mass term, and $M_{1x}$ and $M_{1y}$ are curvature terms. $V_x$ and $V_y$ are the anisotropic couplings between the pseudospin components and $\alpha$ is the Rashba strength. Note that we are in the topological non-trivial regime since $\operatorname{sgn}(M_0/M_{1y})>0$.}
        \label{tab:real_params}
    \end{center}
\end{table}
\end{widetext}
$\alpha$ is a material-dependent parameter proportional to the electric field that induces structural asymmetry and represents the strength of the Rashba interaction. The vector of Pauli matrices  $\boldsymbol{s} = (s_x, s_y, s_z)$ acts on the real spin subspace while $\boldsymbol{\sigma} = (\sigma_x, \sigma_y, \sigma_z)$ acts in the pseudospin (or sublattice) subspace.

In order to analyze the energy band splitting and topological properties, in this section the Rashba coefficient is set to $\alpha = 0.2~ \text{\angstr}^{-1}$ and we choose the model parameters indicated in Table \ref{tab:real_params}. However, in the next section we show the system's conductance response as this Rashba SOC strength is varied within a suitable parameter range. 

Note that the Rashba coupling is anisotropic due to the inherent anisotropy of the velocity operator. Its form interpolates between momentum-independent velocity, characteristic of graphene-like systems, and the linear correction associated with the spin-orbit interaction in conventional two-dimensional electron gases.\\

\subsection{Dispersion relation of the edge states in the presence of RSOC} \label{sec:Dispersion_edge_states}

To study the influence of RSOC on the edge states spectrum, we treat the Rashba term as a perturbation of the spin-independent Hamiltonian $H_{\rm SD}^{4 \times 4} = s_0 \otimes H_0 (\boldsymbol{k})$ in Eq.  \eqref{eq:ham_SD}.
Since this Hamiltonian acts trivially in spin space, its eigenstates are doubly degenerate in spin. In particular, each edge state solution $| \psi_\sigma \rangle$ of the spinless Hamiltonian $H_0$, associated with energy $E_\sigma$ (with $\sigma_{\pm}$), corresponds to two opposite spin-polarized eigenstates of the full Hamiltonian,
\begin{equation}
    \left| \Psi_{\sigma}^{s} \right\rangle \equiv 
    \left| s_z \right\rangle \otimes \left| \psi_{\sigma} \right\rangle,~~ \text{for }s = \uparrow, \downarrow, ~  \sigma = \pm,
\end{equation}
$\left| s_z \right\rangle$ are the eigenstates of the $s_z$ operator in the spin subspace and $\left| \psi_{\pm} \right\rangle$ refers to the edge states in the spinless system given by Eq.~\eqref{eq:edge _states_wf}. For each fixed $\sigma$, the pair $ \lbrace | \Psi_{\sigma}^{\uparrow}\rangle, | \Psi_{\sigma}^{\downarrow}\rangle \rbrace$ spans a two-dimensional degenerate subspace, in which we apply degenerate perturbation theory.

Note that the index $\sigma = \pm$, which labels the edge-branches in the absence of Rashba SOC, is introduced here as a band label. This usage differs from the earlier one, $\boldsymbol{\sigma} = (\sigma_x, \sigma_y, \sigma_z)$, which refers to the pseudoespin degree of freedom associated with the sublattice space in the Hamiltonian. The connection between these two interpretations lies in the pseudospin character of the unperturbed edge states. Specifically, the branch with energy 
$E_+$ is localized on one edge of the sample and is predominantly associated with the B-sublattice pseudospin component, while the $E_-$ branch is localized on the opposite edge and corresponds mainly to the A-sublattice pseudospin. Thus, although $\sigma = \pm$ is used as a band index here, it also indirectly reflects the underlying pseudospin polarization of the edge states in the absence of Rashba coupling.\\

Since the Rashba term,  $U_{\rm RSOC}$, acts only in spin space via $s_x$ or $s_y$, it couples states of opposite spins. As a result, within the subspace $\lbrace |\Psi^{\uparrow}_{\sigma} \rangle , |\Psi^{\downarrow}_{\sigma} \rangle \rbrace$, the RSOC perturbation is given by a $2\times 2$ Hermitian matrix with vanishing diagonal terms,
\begin{equation}
    \langle \Psi^{s}_{\sigma} | \alpha U_{\mathrm{RSOC}}| \Psi^{s'}_{\sigma}\rangle = \begin{pmatrix}
        0 & \chi_{\sigma} \\ \chi_{\sigma}^*&0
    \end{pmatrix},
    \label{eq:deg_perturb}
\end{equation}
which encodes that the pseudospin structure is approximately preserved. Writing $\chi_{\sigma} = |\chi_{\sigma}| \ e^{i \eta_{\sigma}}$ and diagonalizing this matrix, the perturbed eigenenergies are shifted by $\pm |\chi_{\sigma}|$ relative to $E_{\sigma}$ and the corresponding perturbed eigenstates become a superposition of the unperturbed ones, modulated by the phase $\eta_{\sigma}$, 
\begin{align} 
& \left| \  \Psi_{\pm}^{\uparrow} \right\rangle_{\alpha} = \frac{1}{\sqrt{2}} \left( \left| \Psi_{\pm}^{\uparrow} \right\rangle - e^{i \eta_{\pm}} \left| \Psi_{\pm}^{\downarrow} \right\rangle \right) , \\
&\left| \Psi_{\pm}^{\downarrow} \right\rangle_{\alpha} = \frac{1}{\sqrt{2}} \left( \left| \Psi_{\pm}^{\uparrow} \right\rangle + e^{i \eta_{\pm}} \left| \Psi_{\pm}^{\downarrow} \right\rangle \right). 
\end{align}
Note that if $\chi_{\sigma}$ is purely imaginary, the mixing phase becomes  $\eta_{\sigma} = \sigma \pi/2$. In this case, the sign of $\chi_{\sigma}$ can be absorbed into the phase, so when $\chi_{\sigma}$ changes its sign, it effectively exchanges the character of the perturbed eigenstates. \\

From the spinless wavefunctions \eqref{eq:edge _states_wf}, we find $\chi_{\sigma}
= \sigma~ iV_x \text{sgn}(M_{1y} V_y) \alpha k_x$, 
leading to the energy dispersion,
\begin{equation} \label{eq:disp_edgestates}
    E_{\pm} (s) = \pm V_x \text{sgn}(M_{1y} V_y) (k_x^2 \mp \mathbf{s} \alpha k_x).
\end{equation}
Here, $\mathbf{s} = + 1$ for $E_{\pm} (\uparrow)$ and $\mathbf{s}  = -1$ for $E_{\pm} (\downarrow)$. Since $\eta_{\sigma} = \sigma~ \pi/2$, so the mixing of the spin states is independent of the Rashba strength. However, when $k_x$ changes its sign, the perturbed eigenstates exchange their character, and the orientation of the spin polarization along the $s_y$ direction is reversed. 
 
\begin{figure}
    \centering
    \includegraphics[width=0.9\linewidth]{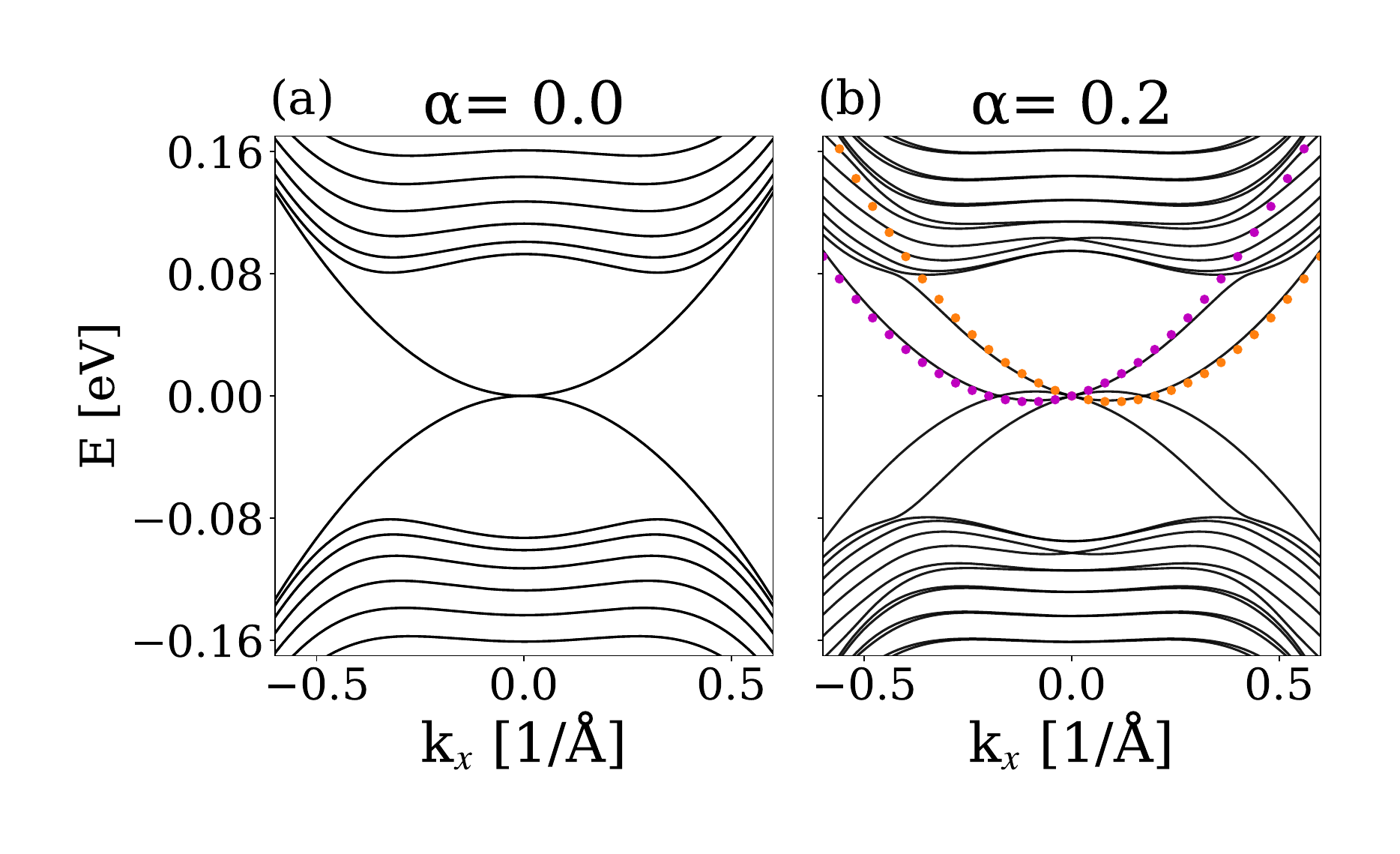}
    \caption{Band structure for a nanoribbon of $66~\angstr$ width and translational symmetry in $x$-direction for the parameters of the model specified in Table \ref{tab:real_params}. The quadratic dispersion relation of the edge states that emerge in the gap of the system is represented without RSOC, $\alpha = 0 ~\text{\angstr}^{-1}$ in panel (a) and for a Rashba strength of $\alpha = 0.2~\text{\angstr}^{-1}$ in panel (b). The solid line corresponds to the numerical results and the dotted lines are the analytical solution given by \eqref{eq:disp_edgestates} for $E_+(s)$, in purple (orange) for $s = \uparrow$ ($s = \downarrow$).} 
    \label{fig:Rashba_bands}
\end{figure}
\begin{widetext}
    \begin{figure*}
        \centering
        \includegraphics[width=0.82\linewidth]{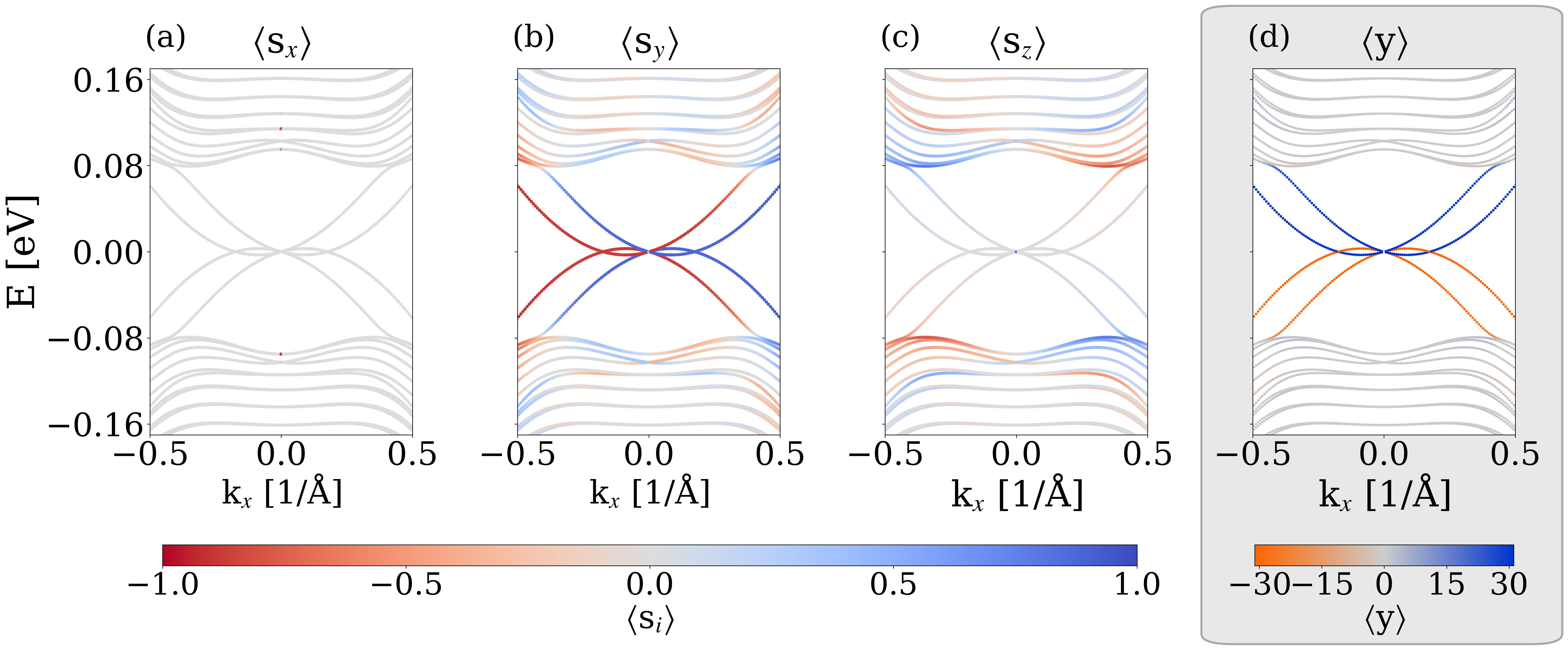}
        \caption{Expected values of the spin operators for the band structure of a nanoribbon of a width of $66~\angstr$, the same considered in Fig.~\ref{fig:Rashba_bands} (b), displaying $\langle s_x\rangle,\langle s_y\rangle$ and $\langle s_z\rangle$ in panels (a)-(c). The color indicates the spin polarization for each spin orientation, red for the spin down and blue for the spin up. In panel (d), we plot the expected value of the $y$ coordinate, which provides information about the spatial location of the edge states. The bands stemming from perturbing $E_+$ lie on the upper edge (shown in dark blue), while those resulting from perturbing $E_-$ are on the lower edge (shown in orange).}
        \label{fig:expected spin-value}
    \end{figure*}
\end{widetext}

In Fig. \ref{fig:Rashba_bands}, we can see the good agreement between this analytical solution, plotted with a dotted line, and the numerical diagonalization in a finite a nanoribbon of width $W= 66~\text{\AA}$, with translation symmetry in the $x$-direction. The purple dotted line represents $E_+ (\uparrow)$ and the orange one, $E_+ (\downarrow)$.
In addition, we plot the expectation value of the spin components in Fig. \ref{fig:expected spin-value}. This figure shows how the RSOC induces spin mixing, thereby lifting the spin degeneracy of the edge states initially aligned along the $s_z$ direction. The central panel of Fig. \ref{fig:expected spin-value}
reveals a strong polarization along the $s_y$ direction, with $\langle s_y \rangle$ reaching values close to $\pm 1$. Notably, the sign of $\langle s_y \rangle$ flips across $k_x = 0$, a behavior that aligns with the fixed value $\eta_{\sigma}$ discussed earlier. The spectrum is completely unpolarized in $s_x$.

In Fig. \ref{fig:expected spin-value} (d), we plot the expected value of the $y$-coordinate in terms of the $k_x$-states, providing a direct link between the spatial localization of the states and their spin polarization under the presence of Rashba SOC. 
Without SOC, the spatial localization of the edge states depends on its energy. The positive energy branch of the band structure is localized on the upper edge, while the negative energy branch resides on the lower edge. When Rashba SOC is introduced, the edge localization is preserved: the bands that come from the unperturbed positive-energy parabola remain localized on the upper edge, even in the energy range where they now appear negative; likewise, the bands originating from the negative-energy parabola stay confined to the lower edge.

For numerical checks, we have considered a lattice regularization of the model using Kwant \cite{GrothKwant}. From the tight-binding perspective, the RSOC introduces spin-flipping hopping terms between sublattices, couplings of $| s \sigma  \rangle$ and $| s' \sigma' \rangle$ components, where the sublattice character can be preserved but not the spin character $s \neq s'$. 

As a consequence, the system is no longer invariant under spin rotations, leading to notable changes in the bulk band structure. Specifically, the RSOC splits each of the two original bands into two with opposite spin orientations. The spin texture exhibits opposite behavior on either side of $k_x=0$, consistent with time-reversal symmetry (see panels (a) and (b) of Fig. \ref{fig:Zak_phase_Sy}).

\subsection{Topological protection} \label{sec:Topogical_protect}

With regard to topological protection, semi-Dirac systems with SOC present a subtle scenario. As we already mentioned, they can host edge states, but these are not chiral and do not extend uniformly along the entire surface of the system \cite{SDpaper_2024}. To characterize the topology of the band structure, we analyze the Berry phase acquired by the Bloch wavefunction of the $n$-th band, $| u_n (\boldsymbol{k}) \rangle$, as it adiabatically evolves around a closed loop in one momentum direction (here $k_y$), while keeping the other component ($k_x$) fixed. In one-dimensional settings, it is often referred to as the Zak phase \cite{Zak1989}.
\begin{equation}
    \mathcal{Z}_n (k_x) = \oint_{-\pi}^{\pi} d k_y \langle u_n (\boldsymbol{k}) | \partial_{k_y} | u_n (\boldsymbol{k}) \rangle
\end{equation}
This quantity can be computed numerically from the overlap of neighboring Bloch states along a discretized path in the $y$-momentum defined by $k_i \in \left[ -\pi/a, \pi/a \right)$ for $i = 1, \dots, N$ with periodic boundary conditions $| u_n (k_N) \rangle = | u_n (k_1) \rangle$ \cite{Fukui2005, Vanderbilt_2018},
{\small
\begin{align}
    \mathcal{Z}_n (k_x) = - \text{Im~ln} \left( \langle u_n (k_1) \right. &| u_n (k_2) \rangle \langle u_n (k_2) | u_n (k_3) \rangle  \dots \\ \nonumber
    & ~ \left. \dots \langle u_n (k_{N-1}) | u_n (k_N) \rangle \right). 
\end{align}
}
When the energy bands are isolated, this approach can be applied band-by-band. The resulting phases directly correspond to polarization charges in 1D systems and signal topological transitions in 2D systems when they cross the $\pi$ value \cite{Soluyanov2011, Vanderbilt_2018}. Moreover, a non-trivial value of $\pi$ in the Zak phase at some fixed $k_x$ implies the presence of zero-energy boundary modes in a 1D system parametrized by that $k_x$. This allows us to capture momentum-dependent topological features in 2D systems, based on reduced dimensional slices \cite{SDpaper_2024}.\\

However, in systems with band crossings or degeneracies, the single-band Berry phase becomes ill-defined due to the ambiguity in the tracking of individual bands. In such cases, one must adopt a multi-band Berry phase, which considers that the occupied states from different bands can mix along the path \cite{Soluyanov2012, Yu2011}.

In the model under consideration, the topological properties can be explored from a single-band perspective across all values of $k_x$, including the bulk band crossing at $(k_x, k_y) = (0,0)$ shown in panels (a) and (b) of Fig. \ref{fig:Zak_phase_Sy}. In fact, even in the absence of RSOC, the system exhibits bands that are fully degenerated in spin, but they are completely decoupled. The Zak phase is found to be $\pi$ for each band and for each spin polarization at $k_x = 0$.  So we obtain a total of four degenerate states at zero-energy \cite{SDpaper_2024}. When the RSOC is considered, it causes the splitting of the bands, giving rise to eight propagating modes at zero-energy. Two types of topologically non-trivial points must be distinguished, with $k_x \neq 0$ and with $k_x=0$. As we explain in the following subsections, in the case $k_x \neq 0$ the bands are intrinsically gapped, while for $k_x=0$ the bands can be decoupled with the correct transformation.

\subsubsection{Topological points at $k_x \neq 0$: Isolated bands and band inversion}

\begin{figure}
    \centering
    \includegraphics[width =\linewidth]{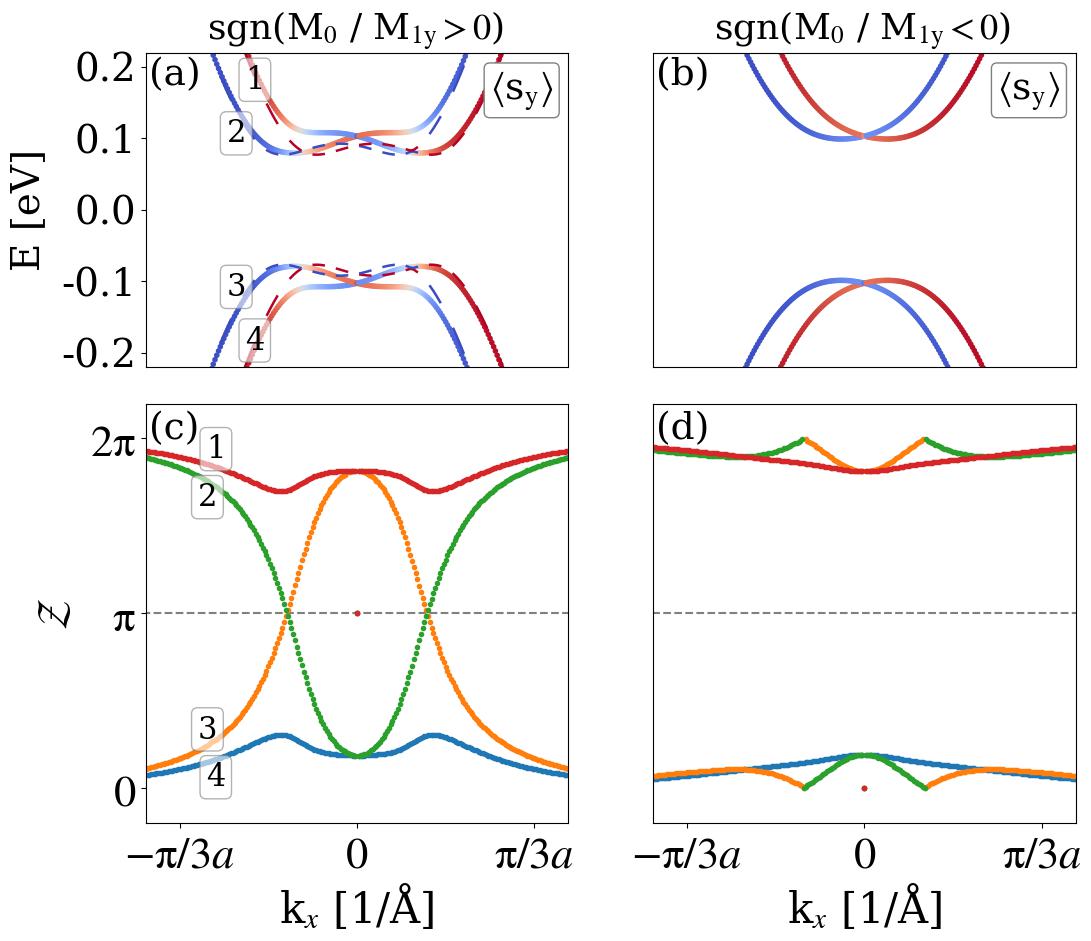}
    \caption{
    Bulk band projections and corresponding single-band Zak phases are shown for both band-inverted ($\rm sgn (M_0/M_{1y}) > 0$) and trivial regimes ($\rm sgn (M_0/M_{1y}) < 0$). Panels (a) and (b) display the bulk band structure along $k_x$ at $k_y = 0$. The expectation value of the spin component $s_y$ is color-coded: red for the down component (-1) and blue for the up component (+1). Panels (c) and (d) show the associated Zak phases for the four bands labeled from 1 to 4.  The non-trivial points with a Zak phase of $\pi$ for bands $n=2,3$ in panel (c) are numerically determined, and they correspond with $k_x=\pm 0.408~\text{\angstr}^{-1}$. Panel (a) includes in dashed lines the case where the $\alpha V_y$ term is removed to highlight the avoided band crossings that the Rashba causes at $k_x= \sqrt{M_0 M_{1x}/(M_{1x}^2+V_x^2)} = \pm 0.324 \angstr^{-1}$.}
    \label{fig:Zak_phase_Sy}
\end{figure}

At $k_x \neq 0$, the projection of the bulk bands along $k_y$ shows four isolated bands, allowing the use of the single-band Zak phase approach. The results are shown in Fig. \ref{fig:Zak_phase_Sy}. Panels (a) and (b) show the projected bulk bands along $k_x$ at $k_y = 0$, for the non-trivial ($\rm sgn( M_0 / M_{1y} )< 0$) and trivial ($\rm sgn( M_0 / M_{1y} ) > 0$) regimes, respectively. The color map encodes the expectation value of $s_y$ operator. The corresponding Zak phases are shown in panels (c) and (d). In the non-trivial regime (panel (c)), we find four non-trivial values of $\pi$, two for band $n=2$ and two for band $n =3$, see Fig.~\ref{fig:Zak_phase_Sy} for the definition of the bands' numbering. 
In panel (a), the system is evaluated with (shown as a continuous line) and without the $\alpha V_y$ contribution (plotted as a dashed line) to the Rashba coupling. This reveals clear avoided crossings between the conduction (and valence) bands in the band-inverted regime. These avoided crossings are a direct signature of spin mixing induced by Rashba SOC and underscore the non-trivial topology of the system. To see this inverted character, we can check the complex-value of the perturbation between degenerate states. As in the case of the edge states, the Rashba interaction only mixes spin sectors, it has the form of \eqref{eq:deg_perturb}, and it is also purely imaginary. It is given by,
\begin{multline}
    \chi_{\sigma} = \sigma~i\alpha \big[ (M_0- M_{1x} k_x^2- M_{1y} k_y^2)\\
    \times (M_{1x}k_x + M_{1y} k_y) - V_x^2k_x^3 -V_y²k_y/2\big]~.
\end{multline}
The $k_x$ value at which the inversion of the spinor occurs is $(k_x, k_y) = (\sqrt{M_0M_{1y}/(M_{1x}^2 + V_x^2)}, 0)$, matching  the location of the band crossing avoided by the term $\alpha V_y$. \\

The non-trivial points of the Zak phase appear slightly shifted at $k_x = \pm 0.408 ~\angstr^{-1}$ for a Rashba strength $\alpha = 0.2~\text{\angstr}^{-1}$. The value of these topological points is proportional to the Rashba strength and depends on the model parameters, but as long as sgn$(M_0/M_{1y}) > 0$, the topological points are present.

\subsubsection{Topological points at $k_x = 0$: Symmetry-protected edge states}

At $k_x = 0$, the system exhibits a bulk band crossing between bands $n = 1, 2$ and $n=3,4$. While one could approach  them numerically using a multi-band Wilson loop, in this case the system decouples into two independent SSH-like chains under suitable basis transformation.\\ 

Before introducing the RSOC, the spin-independent system at $k_x=0$ can be mapped onto a generalized SSH model by performing a lattice regularization of the continuum model \eqref{eq:d-vector} via $k_i \rightarrow {1}/{a} \sin (k_ia), ~ k_i^2\rightarrow {1}/{a^2} (1-\cos(k_ia))$ \cite{Shen2012TopologicalInsulators} and rotating the system in the sublattice subspace. 
The rotation is defined by $H_{0} = R H_{\text{gSSH}} R^{\dagger}$ such that
\begin{align} \label{eq:hamgSSH}
    H_{\text{gSSH}} & = \left[ v + w_1  \cos(ka) \right]\sigma_x  + w_2 \sin(ka) \sigma_y~,
\end{align}
where the parameters of the corresponding SSH chain are given by
\begin{subequations} \label{eq:hamgSSH_coeff}
\begin{align} 
    v & = M_0 - 2 M_{1y}/a^2~,\\
    w_1  &= (M_{1y}+V_y)/a~,\\
    w_2 &= (M_{1y}-V_y)/a.  
\end{align}
\end{subequations}
$v$ is the intracell hopping between $A$ and $B$ sublattices, while $w_1$ is the intercell nearest-neighbor hopping between $A$ and $B$ sublattices. 
The $w_2$ term introduces a third neighbor hopping that preserves chiral symmetry. This means that the Hamiltonian does not allow hoppings between the same sublattice and consequently anticommutes with the chiral  unitary operator, related to the sublattice degree of freedom \cite{Shen2010}.\\ 

\par
\begin{figure}
    \centering
    \includegraphics[width=\linewidth]{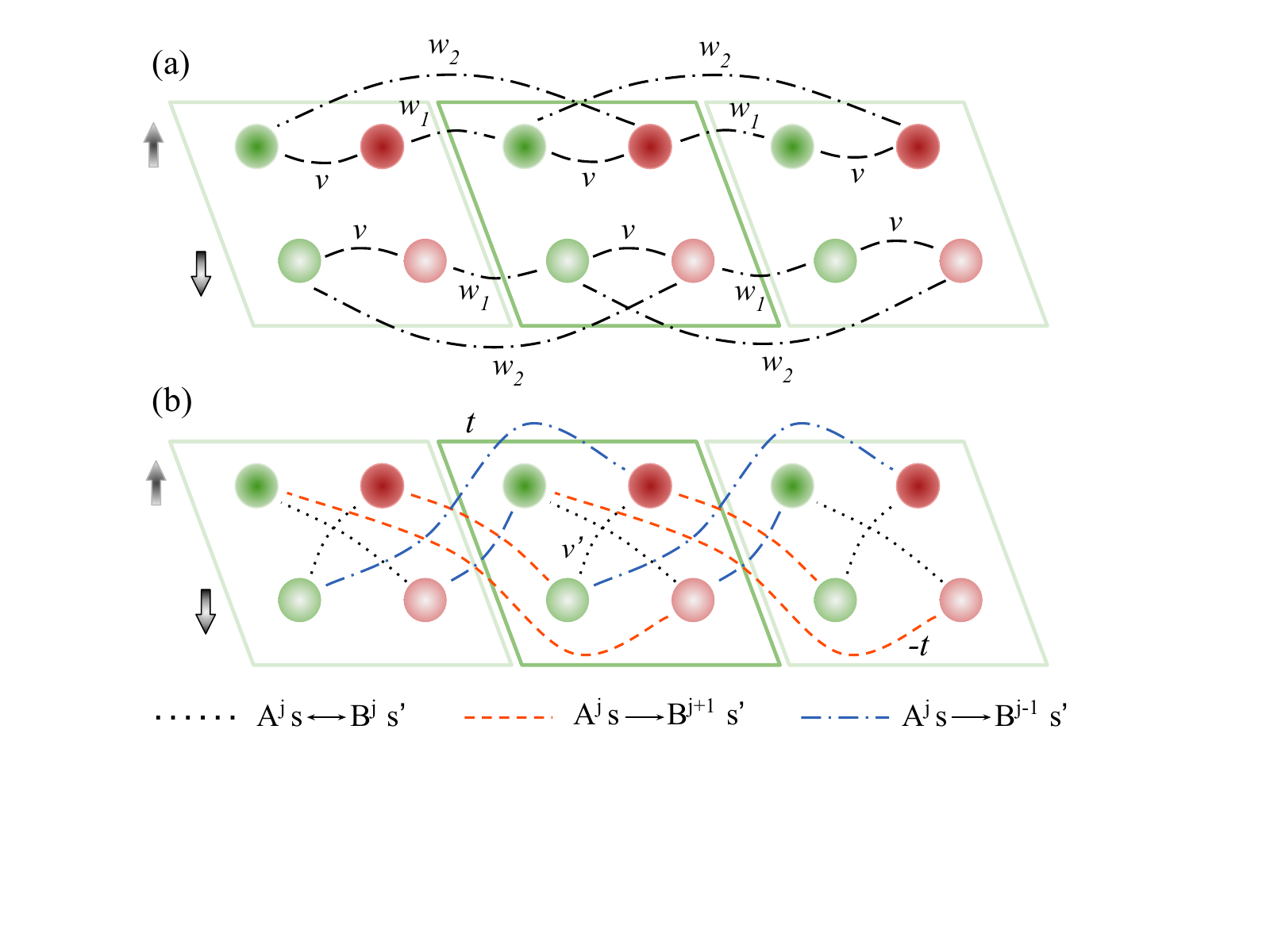}
    \caption{Tight-binding model associated with the one-dimensional system for $k_x = 0$. The color code reflects the sublattice and spin components of the basis: Dark green: sublattice $A$, spin $\uparrow$; dark red: sublattice $B$, spin $\uparrow$; light green: sublattice $A$, spin $\downarrow$; light red: sublattice $B$, spin $\downarrow$. Panel (a) shows the hopping structure in the absence of RSOC interaction. In this limit, the system can be mapped into two independent SSH-like chains with third neighbors for each spin polarization. The system preserves chiral symmetry, and hence no hoppings between different sublattices are present. The values of the hopping terms of the model parameters are given in Eqs.~\eqref{eq:hamgSSH_coeff}. 
    In panel (b) we plot the hopping that the Rashba coupling introduces, i.e. a complex, spin-flip hopping amplitudes between opposite spin states given by Eq.~\eqref{eq:hamgSSH_coeffRSOC}. }
    \label{fig:tbmodel_k_x=0}
\end{figure}

 When the RSOC is considered, the transformed Hamiltonian takes the form, $H_\mathrm{RSOC}^{\text{gSSH}} = (s_0 \otimes R^\dagger) H_\mathrm{RSOC} (s_0 \otimes R)$, which explicitly reads
\begin{equation}
    H_\mathrm{RSOC}^{\text{gSSH}} = \begin{pmatrix}
        0 & \Delta^* & 0 & g \\ \Delta & 0 & g^* & 0 \\ 0 & g & 0 & \Delta^* \\ g^* & 0 & \Delta & 0
    \end{pmatrix},
\end{equation}
where $\Delta = (M_0 - M_{1y} k_y^2- iV_y k_y)$ and $g = \alpha (M_{1y}k_y + iV_y)$. The latter term introduces inter- and  intra-cell hopping terms that flip the spin between sublattices A and B. In this manner, it directly connects the decoupled chains,  mixing A and B subspaces, and modifying the original SSH-like structure into the structure represented in Fig. \ref{fig:tbmodel_k_x=0}. The hopping parameters are now given by
\begin{equation} \label{eq:hamgSSH_coeffRSOC}
    v' =iV_y \alpha~,\quad t = i M_{1y} \alpha.
\end{equation}
Similar generalized SSH models have been studied in the literature \cite{Li2017,Ahmadi2020}.

\par To recover a clear topological picture, we switch to a new basis involving symmetric and antisymmetric spin combinations, $\lbrace | A_{+} \rangle, | B_{+} \rangle, | A_{-} \rangle, | B_{-} \rangle \rbrace$,
\begin{align}
    | A_{\pm} \rangle = \frac{1}{\sqrt{2}} (| A \uparrow \rangle \pm | A \downarrow ), \\
    | B_{\pm} \rangle = \frac{1}{\sqrt{2}} (| B \uparrow \rangle \pm | B \downarrow ).
\end{align}
Via the unitary transformation, $U$,
\begin{equation}
    U = \frac{1}{\sqrt{2}} \begin{pmatrix} 1 & 0 & 1 & 0 \\ 0 & 1 & 0 & 1 \\ 1 & 0 & -1 & 0 \\ 0 & 1 & 0 & -1 \end{pmatrix},
\end{equation}
we obtain the block-diagonal Hamiltonian
\begin{equation}
    H' = \begin{pmatrix}
        0 & h^* & 0 & 0 \\ h & 0 & 0 & 0 \\ 0 & 0 & 0 & h \\ 0 & 0 & h^* & 0
    \end{pmatrix},
\end{equation}
with $h = M_0 - M_{1y}k_y^2 + \alpha M_{1y} k_y + i V_y \left( k_y + \alpha \right)$.\\

\par In this basis, we have obtained again two decoupled SSH-like chains, as in panel (a) of Fig. \ref{fig:tbmodel_k_x=0}. Each of these effective decoupled systems preserves the chiral symmetry and supports zero-energy edge states. The RSOC can then be understood as a modification of the hopping amplitudes, but not of the symmetry structure that preserves the topological protection.

Thus, the number of observed zero-energy edge states matches with the number of non-trivial points, in accordance with the bulk boundary correspondence.\\

 \section{Spin-resolved conductance} \label{sec:spin_resolvedG}

\begin{figure*}
    \centering
    \includegraphics[width = 0.4\linewidth]{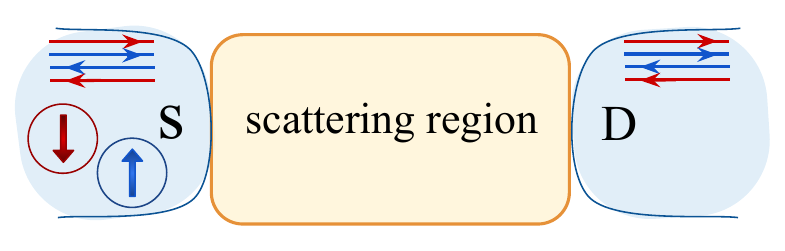}
    \includegraphics[width=0.8\linewidth]{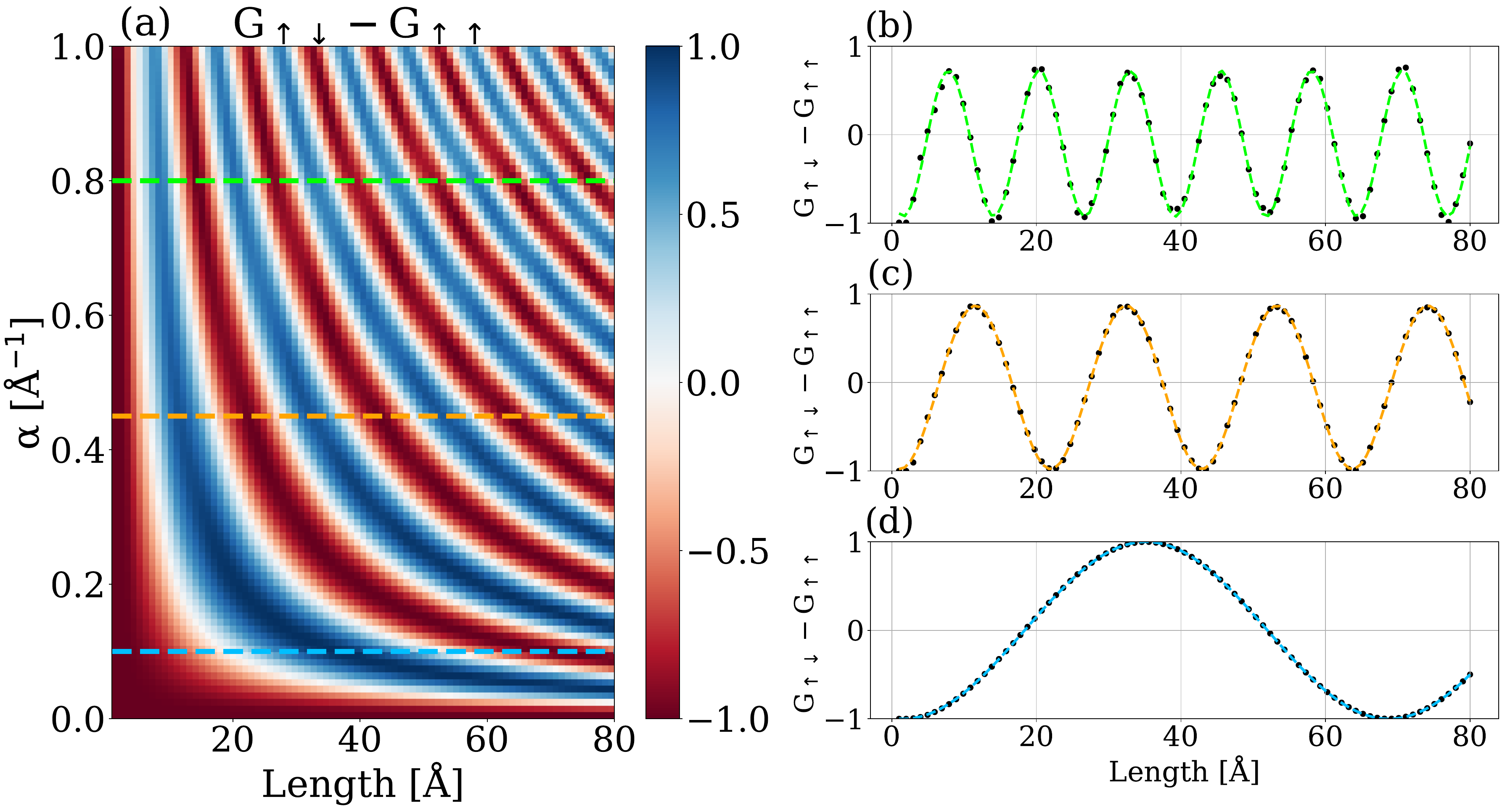}
    \caption{Panel (a) displays the difference $G_{\uparrow\downarrow} -G_{\uparrow \uparrow}$ as a function of the Rashba strength ($y$-axis) and the length of the system ($x$-axis) at a fixed energy of injection $E_F = 0.06~\mathrm{eV}$, a value well inside the first step of conductance. 
    Blue (red) dark areas show regimes in which $G_{\uparrow \downarrow}$ ($G_{\uparrow \uparrow}$) is maximized. In white, the boundary regions indicate a change in behavior. Panels (b)-(d) display $G_{\uparrow \uparrow}-G_{\uparrow \downarrow}$ as a function of the length at three given values of Rashba strength highlighted in panel (a) with color-correspondent-dashed lines. 
    In particular, $\alpha = \{ 0.10, 0.45, 0.80 \} ~\text{\angstr}^{-1}$ in panels (b, c and d), respectively. By fitting the oscillatory dependence to Eq.~\eqref{eq:LR} we obtain $L_R = \{ 67.21, 20.87, 12.56\} \angstr$, for panels (b, c and d), respectively:. }
    \label{fig:conductance_alphaL}
\end{figure*}
In this section, we perform transport simulations on a two-terminal device within the Landauer-B\"uttiker framework \cite{Landauer-Buttiker}. We consider a sample in the $xy$ plane with two leads along the $x$-direction and a width of $W = 66\angstr$ in the $y$-direction, as in the previous section. The system is modeled with open boundary conditions along the $y$-direction and semi-infinite leads described by the same Hamiltonian of the scattering region, but without RSOC. Furthermore, to minimize additional scattering effects arising from a mismatch of the incoming modes and the modes allowed in the scattering region, we set the same Fermi energies for the leads and the scattering region. Thus, any deviations in transport properties are mainly due to the spin dynamics induced by the RSOC. To analyze the contribution of the edge states, we set the injection energy at $E_F = 0.06~\text{eV}$, within the bulk energy gap. 

\par In addition, we assume that the current flows from left to right, meaning that the left lead injects two propagating modes that are eigenstates of the $s_z$ operator, the ones with positive velocity (see scheme of Fig. \ref{fig:conductance_alphaL}). The scattering region is subjected to an electric field applied in the $z$-direction which induces an effective Rashba magnetic field $\boldsymbol{B}_R$ pointing along $y$-direction, oriented perpendicular to both the motion direction and the applied electric field. The torque 
between the spin polarization of the injected state and $\boldsymbol{B}_R$ induces spin precession, leading to oscillations in the spin texture as the states propagate through the system.  \\

To understand the role of the spin in transport, we analyze the spin-resolved conductance components, $G_{s~s'}$, where $s, s' = \uparrow, \downarrow$. That is, the conductance of a mode injected in a spin state $s$ that leaves the system as a spin state $s'$. The total charge conductance is the sum of all the spin components $G = G_{\uparrow \uparrow} + G_{\uparrow \downarrow} + G_{\downarrow \uparrow} + G_{\downarrow \downarrow}$. 
Thus, in the absence of RSOC ($\alpha = 0$), spin is conserved, and the spin-flipped components $G_{\uparrow \downarrow}$ and $G_{\downarrow \uparrow}$ are zero. The system exhibits a perfectly quantized conductance of two unit steps, as each spin channel contributes independently. However, when $\alpha \neq 0$, spin precession occurs due to the hybridization of the states in the scattering region. As a result, the spin-flipped components $G_{ss'}$ for $s\neq s'$, are nonzero, and the originally quantized values of the components $G_{ss}$ decrease, since part of the injected current is now scattered into the opposite spin channel. The precession frequency and amplitude depend on several parameters, including the external electric field (which modifies the RSOC strength), the Fermi energy of the injected states and the length of the system, which determines how much spin rotation occurs before reaching the drain reservoir. By varying these parameters, we can tune the spin-flip probability.

\par To illustrate this, Fig. \ref{fig:conductance_alphaL} shows the difference between spin-up and spin-down conductance $G_{\uparrow\downarrow} -G_{\uparrow \uparrow}$ for an initially injected spin-up state, as a function of $\alpha$ and the system length. An analogous result is obtained for an injected spin-down state. 
If the difference is negative, spin-up states are more likely to leave the system as spin-up rather than spin-down. Conversely, a positive difference suggests that spin-up states are more likely to scatter into spin-down states. This gives rise to the emergence of an oscillatory pattern that indicates the presence of a characteristic precession length. This length scale determines the optimal conditions for maximizing the spin-flipping probability. 

\par By numerically fitting the oscillatory dependence, 
\begin{equation} \label{eq:LR}
    G_{\uparrow \downarrow}-G_{\downarrow \uparrow} = A \cos \left( \frac{2 \pi L}{L_R} + \phi\right) + C~,
\end{equation}
the spin precession length $L_R$ can be computed for a fixed Rashba strength. 
For the set of parameters of \ref{tab:real_params}, we get $L_R = 10.94 /(\alpha + 0.08)$, showing that the spin rotation is faster when the Rashba coupling is stronger, as it is expected.\\ 

In order to explore  the robustness of this phenomenon, we introduced into our numerical simulations an Anderson-like disorder term in the scattering region taking the form, $H_{A,j}  = \sum_i \varepsilon_i c_i^\dagger c_i (s_0 \otimes \sigma_j)$
where $\varepsilon_i$ is a random quantity uniformly distributed between $\left[ -\frac{w}{2}, \frac{w}{2}\right]$. Here, $w$ represents the magnitude of the disorder. This onsite random perturbation is included in two different ways, one that preserves the particle-hole symmetry ($\sigma_j = \sigma_z$) and another
that breaks it ($\sigma_j = \sigma_0$). This symmetry analysis helps us to emphasize the role of particle-hole symmetry, even in the presence of RSOC, due to the symmetry-protected edge states at $k_x = 0$.  We consider the same width, $W=66~\angstr$, and a variable length, $L\in [1,80] ~\angstr$ attached, again, to two leads without RSOC.
\begin{figure}
    \centering
    \includegraphics[width=1.1\linewidth]{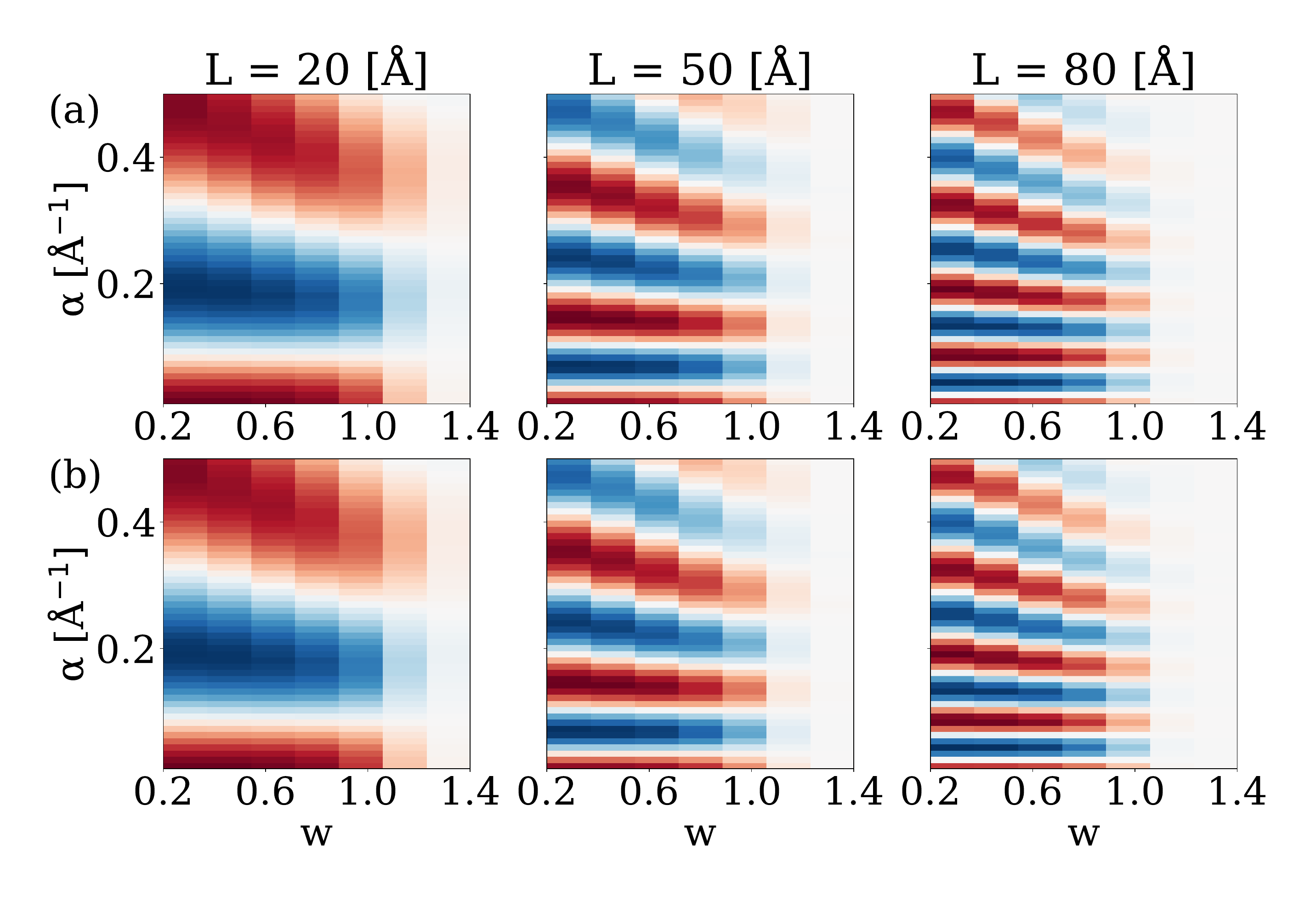}
    \caption{Cut of Fig. \ref{fig:conductance_alphaL} for lengths $20, 50$ and $80~\angstr$ showing the evolution of $G_{\uparrow  \downarrow}-G_{\uparrow \uparrow}$ with the magnitude of disorder $w$. The same color scale of Fig.~\ref{fig:conductance_alphaL}(a) is employed, and we have fixed $E_F = 0.06~\mathrm{eV}$.
    In the upper panels, indicated with (a), we consider that disorder preserves particle-hole symmetry, while in the lower panels, indicated by (b), we show the results for an onsite disorder that breaks all the symmetries. As the disorder grows, we see a regime represented in white color in which the spin is completely randomized due to scattering processes with the disorder present. The data have been averaged over 150 realizations.}
    \label{fig:G_disorder}
\end{figure}

The results are shown in Fig. \ref{fig:G_disorder}. If disorder respects the particle-hole symmetry [panel (a)], the value of the difference between spin channels remains stable over a range of disorder strengths. Therefore, to randomize the spin scattering process, a higher magnitude of disorder is required compared to an Anderson disorder that breaks all symmetries [panel (b)].\\

\section{Conclusions}

We have studied a model of type-I semi-Dirac material that includes both band inversion and Rashba spin-orbit coupling, with a focus on its topological and spin-transport properties. To analyze the bulk-boundary correspondence for this highly anisotropic model, we have used a dimensional reduction approach, wherein the momentum along the direction of the quadratic dispersion is treated as a parameter mapping the system onto an effective one-dimensional model. This reduction enabled us to use the Zak phase as a marker for the non-trivial topological properties.

In the topologically non-trivial regime, we have identified the presence of two different types of zero-energy edge states. For finite momenta $k_x \neq 0$, the emergence of zero modes is linked to the Zak phase of the two bands defining the gap due to an avoided crossing. At $k_x=0$, the system maps onto a non-trivial generalized SSH-ladder model with edge states protected by the particle-hole symmetry. We provide an analytical description of these edge states through a spatially decaying \textit{ansatz} and perturbative treatment of the Rashba term. The Rashba SOC interaction preserves the exponential localization on the upper and lower edges depending on the particle or hole character of the non-perturbed edge states. 

The interplay of these features allows for remarkably clean spin-dependent conductance in two-terminal setups. In particular, we observed oscillations in the difference $G_{\uparrow\downarrow} - G_{\downarrow\uparrow}$, which can be interpreted through a simple spin precession model, which inversely scales with the Rashba SOC strength. Furthermore, the robustness of the spin conductance against disorder highlights the potential practical relevance of the system and the key role of the symmetry-protected edge states.

Given the strong control over spin transport demonstrated in this work, bidimensional type-I semi-Dirac materials with Rashba coupling and band inversion offer promising avenues for spintronics applications. Their robust, tunable spin filtering and spin precession effects could be used in the design of efficient spin transistors, spin-based logic devices, and highly sensitive spin sensors. The intrinsic topological protection of edge modes, together with the resilience against disorder, positions these materials as attractive platforms for developing next-generation, low-dissipation spintronic technologies.

\acknowledgments

This work has been supported by Comunidad de Madrid “Talento Program” (Grant 2019-T1/IND-14088), the Agencia Estatal de Investigación from Spain (Grant PID2022-136285NB-C31/C32) and FEDER/Junta de Castilla y León Research (Grant No. SA106P23). M. G. O. acknowledges financial support from the Consejería de Educación, Junta de Castilla y León, and ERDF/FEDER. 
\bibliographystyle{unsrt} 
\bibliography{AA_Biblio_SD_Rashba,references}

\begin{thebibliography}{10}

\bibitem{Ahn2020}
E.~C. Ahn.
\newblock 2d materials for spintronic devices.
\newblock {\em npj 2D Materials and Applications}, 4(1):17, Jun 2020.

\bibitem{reviewZutic2004}
I.~\ifmmode \check{Z}\else \v{Z}\fi{}uti\ifmmode~\acute{c}\else \'{c}\fi{},
  J.~Fabian, and S.~Das~Sarma.
\newblock Spintronics: Fundamentals and applications.
\newblock {\em Rev. Mod. Phys.}, 76:323--410, Apr 2004.

\bibitem{Manchon2015}
A.~Manchon, H.~C. Koo, J.~Nitta, S.~M. Frolov, and R.~A. Duine.
\newblock New perspectives for rashba spin--orbit coupling.
\newblock {\em Nature Materials}, 14(9):871--882, Sep 2015.

\bibitem{Zollner2025}
K.~Zollner, M.~Kurpas, M.~Gmitra, and J.~Fabian.
\newblock First-principles determination of spin--orbit coupling parameters in
  two-dimensional materials.
\newblock {\em Nature Reviews Physics}, Mar 2025.

\bibitem{Bychkov1984}
Y.~A. Bychkov and E.~I. Rashba.
\newblock Properties of a 2d electron gas with lifted spectral degeneracy.
\newblock {\em P. Zh. Eksp. Teor. Fiz.}, 39:66, 1984.

\bibitem{Dresselhaus1955}
G.~Dresselhaus.
\newblock Spin-orbit coupling effects in zinc blende structures.
\newblock {\em Phys. Rev.}, 100:580, Oct 1955.

\bibitem{Datta_Das_transist}
S.~{Datta} and B.~{Das}.
\newblock {Electronic analog of the electro-optic modulator}.
\newblock {\em Applied Physics Letters}, 56(7):665--667, February 1990.

\bibitem{Nitta1997}
J.~Nitta, T.~Akazaki, H.~Takayanagi, and T.~Enoki.
\newblock Gate control of spin-orbit interaction in an inverted
  i${\mathrm{n}}_{0.53}$g${\mathrm{a}}_{0.47}$as/i${\mathrm{n}}_{0.52}$a${\mathrm{l}}_{0.48}$as
  heterostructure.
\newblock {\em Phys. Rev. Lett.}, 78:1335--1338, Feb 1997.

\bibitem{Schultz1999}
M.~Schultz, F.~Heinrichs, U.~Merkt, T.~Colin, T.~Skauli, and S.~Løvold.
\newblock Rashba spin splitting in a gated hgte quantum well.
\newblock {\em Semicond. Sci. Technol.}, 11:1168, 01 1999.

\bibitem{Carbotte2019}
J.~P. Carbotte, K.~R. Bryenton, and E.~J. Nicol.
\newblock Optical properties of a semi-dirac material.
\newblock {\em Physical Review B}, 2019.

\bibitem{Huang2023}
Y.~Huang and R.~Shen.
\newblock The generation and detection of the spin-valley-polarization in
  semi-dirac materials.
\newblock {\em Physica Scripta}, 2023.

\bibitem{Link2018}
J.~M. Link, B.~N. Narozhny, E.~I. Kiselev, and J.~Schmalian.
\newblock Out-of-bounds hydrodynamics in anisotropic dirac fluids.
\newblock {\em Phys. Rev. Lett.}, 120:196801, May 2018.

\bibitem{Xiong2023}
Q.-Y. Xiong, J.-Y. Ba, H.-J. Duan, M.-X. Deng, Y.-M. Wang, and R.-Q. Wang.
\newblock Optical conductivity and polarization rotation of type-ii semi-dirac
  materials.
\newblock {\em Phys. Rev. B}, 107:155150, Apr 2023.

\bibitem{Chan2023}
W.~J. Chan, L.~K. Ang, and Y.~S. Ang.
\newblock Quantum transport and shot noise in two-dimensional semi-dirac
  system.
\newblock {\em Applied Physics Letters}, 2023.

\bibitem{Dietl2008}
P.~Dietl, F.~Pi\'echon, and G.~Montambaux.
\newblock New magnetic field dependence of landau levels in a graphenelike
  structure.
\newblock {\em Phys. Rev. Lett.}, 100:236405, Jun 2008.

\bibitem{Montambaux2009}
G.~Montambaux, F.~Pi\'echon, J.-N. Fuchs, and M.~O. Goerbig.
\newblock Merging of dirac points in a two-dimensional crystal.
\newblock {\em Phys. Rev. B}, 80:153412, Oct 2009.

\bibitem{Delplace2010}
P.~Delplace and G.~Montambaux.
\newblock Semi-dirac point in the hofstadter spectrum.
\newblock {\em Phys. Rev. B}, 82:035438, Jul 2010.

\bibitem{Goerbig2008}
M.~O. Goerbig, J.-N. Fuchs, G.~Montambaux, and F.~Pi\'echon.
\newblock Tilted anisotropic dirac cones in quinoid-type graphene and
  $\ensuremath{\alpha}\text{\ensuremath{-}}{(\text{BEDT-TTF})}_{2}{\text{i}}_{3}$.
\newblock {\em Phys. Rev. B}, 78:045415, Jul 2008.

\bibitem{Pardo2009}
V.~Pardo and W.~E. Pickett.
\newblock Half-metallic semi-dirac-point generated by quantum confinement in
  ${\mathrm{tio}}_{2}/{\mathrm{vo}}_{2}$ nanostructures.
\newblock {\em Phys. Rev. Lett.}, 102:166803, Apr 2009.

\bibitem{Banerjee2009}
S.~Banerjee, R.~R.~P. Singh, V.~Pardo, and W.~E. Pickett.
\newblock Tight-binding modeling and low-energy behavior of the semi-dirac
  point.
\newblock {\em Phys. Rev. Lett.}, 103:016402, Jul 2009.

\bibitem{Huang2015}
H.~Huang, Z.~Liu, H.~Zhang, W.~Duan, and D.~Vanderbilt.
\newblock Emergence of a chern-insulating state from a semi-dirac dispersion.
\newblock {\em Phys. Rev. B}, 92:161115, Oct 2015.

\bibitem{Katayama2006}
S.~Katayama, A.~Kobayashi, and Y.~Suzumura.
\newblock Electric conductivity of the zero-gap semiconducting state in
  $\alpha$-({BEDT-TTF})$_2${I}$_3$ salt.
\newblock {\em J. Phys. Soc. Jap.}, 75:023708, 2006.

\bibitem{Zhong2016}
C.~Zhong, Y.~Chen, Y.~Xie, Y.-Y. Sun, and S.~Zhang.
\newblock Semi-dirac semimetal in silicene oxide.
\newblock {\em Physical chemistry chemical physics : PCCP}, 19 5:3820--3825,
  2016.

\bibitem{CastellanosGomez2014}
A.~Castellanos-Gomez, L.~Vicarelli, E.~Prada, J.~O. Island, K.~L.
  Narasimha-Acharya, S.~I. Blanter, D.~J. Groenendijk, M.~Buscema, G.~A.
  Steele, J.~V. Alvarez, H.~W. Zandbergen, J.~J. Palacios, and H.~S.~J. van~der
  Zant.
\newblock Isolation and characterization of few-layer black phosphorus.
\newblock {\em 2D Materials}, 1(2):025001, jun 2014.

\bibitem{Rodin2014}
A.~S. Rodin, A.~Carvalho, and A.~H. Castro~Neto.
\newblock Strain-induced gap modification in black phosphorus.
\newblock {\em Phys. Rev. Lett.}, 112:176801, May 2014.

\bibitem{PeraltaMireles2024}
M.~Peralta, D.~A. Freire, R.~Gonz\'alez-Hern\'andez, and F.~Mireles.
\newblock Spin-orbit coupling effects in single-layer phosphorene.
\newblock {\em Phys. Rev. B}, 110:085404, Aug 2024.

\bibitem{Zhai2011}
F.~Zhai, P.~Mu, and K.~Chang.
\newblock Energy spectrum of dirac electrons on the surface of a topological
  insulator modulated by a spiral magnetization superlattice.
\newblock {\em Phys. Rev. B}, 83:195402, May 2011.

\bibitem{Li2011}
Q.~Li, P.~Ghosh, J.~D. Sau, S.~Tewari, and S.~Das~Sarma.
\newblock Anisotropic surface transport in topological insulators in proximity
  to a helical spin density wave.
\newblock {\em Phys. Rev. B}, 83:085110, Feb 2011.

\bibitem{Smith2024}
M.~Smith, Victor~L. Quito, A.~A. Burkov, P.~P. Orth, and I.~Martin.
\newblock Theory for ${\mathrm{cd}}_{3}{\mathrm{as}}_{2}$ thin films in the
  presence of magnetic fields.
\newblock {\em Phys. Rev. B}, 109:155136, Apr 2024.

\bibitem{Wu2014}
Y.~Wu.
\newblock A semi-dirac point and an electromagnetic topological transition in a
  dielectric photonic crystal.
\newblock {\em Opt. Express}, 22(2):1906--1917, Jan 2014.

\bibitem{PhysRevX.9.031010}
M.~Mili\ifmmode \acute{c}\else \'{c}\fi{}evi\ifmmode~\acute{c}\else \'{c}\fi{},
  G.~Montambaux, T.~Ozawa, O.~Jamadi, B.~Real, I.~Sagnes, A.~Lema\^{\i}tre,
  L.~Le~Gratiet, A.~Harouri, J.~Bloch, and A.~Amo.
\newblock Type-iii and tilted dirac cones emerging from flat bands in photonic
  orbital graphene.
\newblock {\em Phys. Rev. X}, 9:031010, Jul 2019.

\bibitem{Bellec2013}
M.~Bellec, U.~Kuhl, G.~Montambaux, and F.~Mortessagne.
\newblock Topological transition of dirac points in a microwave experiment.
\newblock {\em Phys. Rev. Lett.}, 110:033902, Jan 2013.

\bibitem{Real2020}
B.~Real, O.~Jamadi, M.~Mili\ifmmode \acute{c}\else
  \'{c}\fi{}evi\ifmmode~\acute{c}\else \'{c}\fi{}, N.~Pernet, P.~St-Jean,
  T.~Ozawa, G.~Montambaux, I.~Sagnes, A.~Lema\^{\i}tre, L.~Le~Gratiet,
  A.~Harouri, S.~Ravets, J.~Bloch, and A.~Amo.
\newblock Semi-dirac transport and anisotropic localization in polariton
  honeycomb lattices.
\newblock {\em Phys. Rev. Lett.}, 125:186601, Oct 2020.

\bibitem{Uryszek2019}
M.~D. Uryszek, E.~Christou, A.~Jaefari, F.~Kr\"uger, and B.~Uchoa.
\newblock Quantum criticality of semi-dirac fermions in $2+1$ dimensions.
\newblock {\em Phys. Rev. B}, 100:155101, Oct 2019.

\bibitem{Kotov2021}
V.~N. Kotov, B.~Uchoa, and O.~P. Sushkov.
\newblock Coulomb interactions and renormalization of semi-dirac fermions near
  a topological lifshitz transition.
\newblock {\em Phys. Rev. B}, 103:045403, Jan 2021.

\bibitem{Mohanta2021}
N.~Mohanta, J.~M. Ok, J.~Zhang, H.~Miao, E.~Dagotto, H.~N. Lee, and S.~Okamoto.
\newblock Semi-dirac and weyl fermions in transition metal oxides.
\newblock {\em Phys. Rev. B}, 104:235121, Dec 2021.

\bibitem{Mondal2022}
S.~Mondal and S.~Basu.
\newblock Vanishing of the quantum spin hall phase in a semi-dirac kane-mele
  model.
\newblock {\em Phys. Rev. B}, 105:235409, Jun 2022.

\bibitem{Saha2016}
K.~Saha.
\newblock Photoinduced chern insulating states in semi-dirac materials.
\newblock {\em Phys. Rev. B}, 94:081103, Aug 2016.

\bibitem{PRX2024}
Y.~Shao, S.~Moon, A.~N. Rudenko, J.~Wang, J.~Herzog-Arbeitman, M.~Ozerov,
  D.~Graf, Z.~Sun, R.~Queiroz, S.~H. Lee, Y.~Zhu, Z.~Mao, M.~I. Katsnelson,
  B.~A. Bernevig, D.~Smirnov, A.~J. Millis, and D.~N. Basov.
\newblock Semi-dirac fermions in a topological metal.
\newblock {\em Phys. Rev. X}, 14:041057, Dec 2024.

\bibitem{Zhang2017}
H.~Zhang, Y.~Xie, Z.~Zhang, C.~Zhong, Y.~Li, Z.~Chen, and Y.~Chen.
\newblock Dirac nodal lines and tilted semi-dirac cones coexisting in a striped
  boron sheet.
\newblock {\em The Journal of Physical Chemistry Letters}, 8(8):1707--1713, Apr
  2017.

\bibitem{SDpaper_2024}
M.~García Olmos, Y.~Baba, M.~Amado, and R.~A. Molina.
\newblock Zero momentum topological insulator in 2d semi-dirac materials.
\newblock {\em Journal of Physics: Materials}, 7(4):045008, oct 2024.

\bibitem{Vanderbilt_2018}
D.~Vanderbilt.
\newblock {\em Berry Phases in Electronic Structure Theory: Electric
  Polarization, Orbital Magnetization and Topological Insulators}.
\newblock Cambridge University Press, 2018.

\bibitem{QSHEreview}
J.~Maciejko, T.~L. Hughes, and S.-C. Zhang.
\newblock The quantum spin hall effect.
\newblock {\em Annual Review of Condensed Matter Physics}, 2(1):31--53, 2011.

\bibitem{Winkler2003}
R.~Winkler.
\newblock {\em Spin-orbit Coupling Effects in Two-Dimensional Electron and Hole
  Systems}.
\newblock Springer Tracts in Modern Physics. Springer Berlin Heidelberg, 2003.

\bibitem{GrothKwant}
C.~Groth, M.~Wimmer, A.~Akhmerov, and X.~Waintal.
\newblock Kwant: a software package for quantum transport.
\newblock {\em New Journal of Physics}, 16, 2013.

\bibitem{Zak1989}
J.~Zak.
\newblock Berry's phase for energy bands in solids.
\newblock {\em Phys. Rev. Lett.}, 62:2747--2750, Jun 1989.

\bibitem{Fukui2005}
T.~Fukui, Y.~Hatsugai, and H.~Suzuki.
\newblock Chern numbers in discretized brillouin zone: Efficient method of
  computing (spin) hall conductances.
\newblock {\em Journal of the Physical Society of Japan}, 74(6):1674--1677,
  2005.

\bibitem{Soluyanov2011}
A.~A. Soluyanov and D.~Vanderbilt.
\newblock Computing topological invariants without inversion symmetry.
\newblock {\em Phys. Rev. B}, 83:235401, Jun 2011.

\bibitem{Soluyanov2012}
A.~A. Soluyanov and D.~Vanderbilt.
\newblock Smooth gauge for topological insulators.
\newblock {\em Phys. Rev. B}, 85:115415, Mar 2012.

\bibitem{Yu2011}
R.~Yu, X.~L. Qi, B.~A. Bernevig, Z.~Fang, and X.~Dai.
\newblock Equivalent expression of ${\mathbb{z}}_{2}$ topological invariant for
  band insulators using the non-abelian berry connection.
\newblock {\em Phys. Rev. B}, 84:075119, Aug 2011.

\bibitem{Shen2012TopologicalInsulators}
S.-Q. Shen.
\newblock {\em {Topological Insulators}}, volume 174 of {\em Springer Series in
  Solid-State Sciences}.
\newblock Springer, Berlin, 2012.

\bibitem{Shen2010}
W.-Y. Shan, H.-Z. Lu, and S.-Q. Shen.
\newblock {Effective continuous model for surface states and thin films of
  three-dimensional topological insulators}.
\newblock {\em New J. Phys.}, 12(4):043048, apr 2010.

\bibitem{Li2017}
C.~Li, S.~Lin, G.~Zhang, and Z.~Song.
\newblock Topological nodal points in two coupled su-schrieffer-heeger chains.
\newblock {\em Phys. Rev. B}, 96:125418, Sep 2017.

\bibitem{Ahmadi2020}
N.~Ahmadi, J.~Abouie, and D.~Baeriswyl.
\newblock Topological and nontopological features of generalized
  su-schrieffer-heeger models.
\newblock {\em Phys. Rev. B}, 101:195117, May 2020.

\bibitem{Landauer-Buttiker}
M.~B\"uttiker.
\newblock Four-terminal phase-coherent conductance.
\newblock {\em Phys. Rev. Lett.}, 57:1761--1764, Oct 1986.

\end{thebibliography}

\end{document}